\documentclass[aps,
reprint,
frontmatterverbose,
 amsmath,amssymb,
 superscriptaddress,
 prl
]{revtex4-1}
\usepackage[table,xcdraw]{xcolor}
\usepackage{graphicx,setspace}
\usepackage{dcolumn}
\usepackage{bm}
\usepackage[mathlines]{lineno}
\usepackage[colorlinks=true,citecolor=blue,linkcolor=blue,anchorcolor=blue,urlcolor=blue]{hyperref}
\usepackage{mathrsfs}
\usepackage[normalem]{ulem}
\usepackage[utf8]{inputenc}
\usepackage{tensor}
\usepackage{physics}
\usepackage{blindtext}
\usepackage{slashed}
\usepackage{tikz}
\usetikzlibrary{decorations.pathmorphing}
\usepackage{cancel}
\allowdisplaybreaks[4]

\usepackage{graphicx}
\usepackage{color}
\usepackage{hyperref}

\begin{document}

\title{Finite-cutoff Holographic Thermodynamics}
\author{Ming Zhang}
\email{mingzhang@jxnu.edu.cn }
\affiliation{Department of Physics and Astronomy, University of Waterloo,  200 University Ave W, Waterloo, Ontario N2L 3G1, Canada}
\affiliation{Perimeter Institute for Theoretical Physics, 31 Caroline St. N., Waterloo, Ontario N2L 2Y5, Canada}
\affiliation{Department of Physics, Jiangxi Normal University, Nanchang 330022, China}
\author{Wen-Di Tan}
\email{wendi.tan@ulb.be}
\affiliation{Physique Théorique et Mathématique, Université libre de Bruxelles and International Solvay Institutes, Campus Plaine C.P. 231, B-1050 Bruxelles, Belgium}
\author{Mengqi Lu}
\email{m64lu@uwaterloo.ca}
\affiliation{Department of Physics and Astronomy, University of Waterloo,  200 University Ave W, Waterloo, Ontario N2L 3G1, Canada}
\author{Dyuman Bhattacharya}
\email{d7bhatta@uwaterloo.ca}
\affiliation{Department of Physics and Astronomy, University of Waterloo,  200 University Ave W, Waterloo, Ontario N2L 3G1, Canada}
\author{Jiayue Yang}
\email{j43yang@uwaterloo.ca}
\affiliation{Department of Applied Mathematics, University of Waterloo, 200 University Ave W, Waterloo, Ontario N2L 3G1, Canada}
\affiliation{Institute for Quantum Computing, University of Waterloo, 200 University Ave W, Waterloo, Ontario N2L 3G1, Canada}
\affiliation{Perimeter Institute for Theoretical Physics, 31 Caroline St. N., Waterloo, Ontario N2L 2Y5, Canada}
\affiliation{Department of Physics and Astronomy, University of Waterloo,  200 University Ave W, Waterloo, Ontario N2L 3G1, Canada}
\author{Robert B. Mann}
\email{rbmann@uwaterloo.ca}
\affiliation{Department of Physics and Astronomy, University of Waterloo,  200 University Ave W, Waterloo, Ontario N2L 3G1, Canada}
\affiliation{Perimeter Institute for Theoretical Physics, 31 Caroline St. N., Waterloo, Ontario N2L 2Y5, Canada}
\affiliation{Department of Applied Mathematics, University of Waterloo, 200 University Ave W, Waterloo, Ontario N2L 3G1, Canada}
\affiliation{Institute for Quantum Computing, University of Waterloo, 200 University Ave W, Waterloo, Ontario N2L 3G1, Canada}
%\date{\today}

\begin{abstract}

We develop a framework for holographic thermodynamics in finite-cutoff holography, extending the anti-de Sitter/conformal field theory (AdS/CFT) correspondence to incorporate a finite radial cutoff in the bulk and a $T^2$-deformed CFT on the boundary. We formulate the first laws  of thermodynamics for a Schwarzschild-AdS (SAdS) black hole with a Dirichlet cutoff on the quasilocal boundary and its dual deformed CFT, introducing the deformation parameter as a thermodynamic variable. The holographic Euler relation for the deformed CFT  and its equation of state  are derived, alongside the Smarr relation for the bulk. We show that the Rupert teardrop coexistence curve defines a phase space island where deformation flow alters states, with up to three deformed CFTs or cut-off SAdS sharing a same phase transition temperature, one matching the seed CFT or original SAdS. These results offer insights into gravitational thermodynamics with boundary constraints and quantum gravity in finite spacetime regions.

\end{abstract}

\maketitle

\newpage

\textit{Introduction---}The equivalence principle and the relativistic paradigm  introduce a fundamentally new geometric framework for understanding gravity at macroscopic and cosmological scales. Contradistinctively, the development of the Standard Model has demonstrated that the electromagnetic, weak, and strong nuclear forces are effectively described by gauge theories grounded in the principle of gauge symmetry \cite{Bousso:2002ju,Susskind:1994vu}. Another foundational concept—the holographic principle \cite{tHooft:1993dmi, Susskind:1994vu}, states that gravitational physics is dual to quantum physics. This principle is embodied by the anti-de Sitter/conformal field theory (AdS/CFT) correspondence \cite{ Maldacena:1997re, Gubser:1998bc, Witten:1998qj, deHaro:2000wj,Strominger:1996sh}, which proposes that a gravitational theory in a $(d+1)$-dimensional AdS spacetime (e.g., an AdS black hole) is equivalent to a strongly coupled gauge theory (at finite temperature in the case of a black hole) defined on its $d$-dimensional conformal boundary. In this duality, the generating functionals of gravitational and gauge theories coincide; bulk fields correspond directly to gauge-invariant operators on the boundary, and the source terms for these dual operators are determined by the leading boundary behavior of bulk fields near the boundary \cite{hawking1983thermodynamics, Witten:1998zw,natsuume2015ads, Papadimitriou:2004ap}.

The thermodynamics of AdS black holes has been extensively studied following the discovery of the Hawking-Page phase transition in Schwarzschild-AdS black holes \cite{Hawking:1982dh}. Phase structures analogous to the van der Waals liquid–gas system have been identified for charged Reissner-Nordström AdS black holes, exhibiting first-order phase transitions and critical phenomena \cite{Chamblin:1999tk,Chamblin:1999hg,Kubiznak:2012wp}. Recently, significant advances have been achieved in AdS black hole thermodynamics by treating the cosmological constant as a variable, resulting in phenomena such as fluid/superfluid phase transitions \cite{Hennigar:2016xwd}, black hole microstructures \cite{Wei:2019uqg}, and more \cite{Kastor:2009wy,Dolan:2010ha,Dolan:2011xt,Lu:2012xu,Xiao:2023lap,Frassino:2022zaz}. The study of thermodynamics within this extended phase space has developed into a subdiscipline known as black hole chemistry \cite{Kubiznak:2016qmn}.

The Hawking-Page phase transition of the Schwarzschild-AdS black hole was holographically interpreted as the confinement-deconfinement phase transition of the quark-gluon plasma in the boundary CFT \cite{Witten:1998zw}, 
and subsequently understood as a solid-liquid transition in the context of black hole chemistry \cite{Kubiznak:2014zwa}.
The holographic interpretation of black hole chemistry (i.e., holographic thermodynamics) has recently garnered significantly  more  attention. Exploratory studies \cite{Johnson:2014yja,Dolan:2014cja,Kastor:2014dra,Zhang:2014uoa,Dolan:2016jjc} indicate that the thermodynamic pressure $P$ and its conjugate volume $V$ in the bulk correspond to the central charge $c$ and its conjugate chemical potential $\mu$ on the boundary. Given that the background metric for the CFT is obtained through conformal completion of the  AdS bulk, there is flexibility in rescaling the spatial radius of the boundary CFT. Specifically, the boundary radius $R$ may either match the AdS radius $\ell$ directly \cite{Karch:2015rpa}, or differ by a conformal factor $\omega=R/\ell$ \cite{Gubser:1998bc,Witten:1998qj}. 
 Rather than varying Newton's constant \cite{Cong:2021fnf} or fixing the spatial volume of the boundary CFT \cite{Zeyuan:2021uol}, a recent study has treated the conformal rescaling factor as an independent thermodynamic parameter \cite{Ahmed:2023snm}. This approach established an exact duality between the first law of extended black hole thermodynamics in the bulk and the first law in the dual CFT, ensuring independent variations of the central charge and spatial volume in the dual theory.

Modern holographic thermodynamics  fundamentally builds upon the AdS/CFT correspondence, where the CFT is ultraviolet (UV) complete. In this setup, an external observer measures properties of gravity contained within an AdS enclosure. Key questions arise: how do we define a quantum gravity model with particular boundary constraints, and can holography provide a meaningful framework in this context? How can we capture the perspective of observers localized in a region of AdS space and make sense of gravity beyond the region's boundary yet well below the AdS scale, i.e., how can we define  bulk observables at some finite distance? Developing holographic thermodynamics from this perspective will deepen our understanding of the holographic principle and provide insights to these questions.

To deal with specified spatial boundary conditions, the framework of the holographic Wilsonian renormalization group (RG) was developed \cite{deBoer:1999tgo,deHaro:2000vlm}, where radial flow in the bulk geometry corresponds to RG flow on the boundary theory, and a radial cutoff in the bulk is interpreted as a Wilsonian cutoff in the dual field theory \cite{Verlinde:1999xm,Freidel:2008sh,Faulkner:2010jy,Heemskerk:2010hk,Papadimitriou:2016yit,Lee:2013dln}. The $T\bar{T}$-deformed two-dimensional CFT, which possesses a large central charge and represents the simplest non-trivial integrable deformation of two-dimensional CFTs \cite{Zamolodchikov:2004ce,Dubovsky:2012wk,Smirnov:2016lqw,Cavaglia:2016oda,Cardy:2018sdv}, admits an exact holographic RG flow equation that fully describes the effective UV cutoff in terms of an irrelevant deformation parameter. Generalization of the $T\bar{T}$ deformation to a higher dimensional CFT has been explored in \cite{Cardy:2018sdv,Bonelli:2018kik,Ferko:2024yua}; specifically, a holographic formulation of a $T^2$ deformation has been proposed \cite{Taylor:2018xcy,Hartman:2018tkw}.

Generalizing the AdS/CFT duality to an effective quantum field theory with a UV cutoff remains a longstanding challenge. Finite-cutoff holography posits  that an effective theory with a UV cutoff corresponds to a gravity dual with a radial cutoff \cite{McGough:2016lol}. At present, perfect agreement has been established  between the energy in $T\bar{T}$ and $T^2$ deformed CFTs  and the quasilocal mass \cite{Brown:1994gs} of its AdS counterparts at successive finite cutoff \cite{McGough:2016lol,Hartman:2018tkw,Kraus:2018xrn,Taylor:2018xcy}. Both of these   quantities can be derived from the stress-energy tensor.  A natural question then arises: how should we formulate holographic thermodynamics adapted to  finite-cutoff holography? In particular, we ask how the standard AdS/CFT holographic dictionary can be refined in   finite-cutoff holography to reveal potential connections between black hole chemistry and the thermodynamics of the dual effective field theory.

Here we take the first step in extending holographic black hole chemistry in the traditional AdS/CFT setting to finite-cutoff holography. We construct the first laws of quasilocal thermodynamics 
\cite{Brown:1994gs} 
(and their integral forms) for both the bulk and the boundary in this context. We also demonstrate how the  deformation affects phase transitions on both sides. To be specific, we will consider the $T^2$ deformation case. Holographic thermodynamics for the $T\bar{T}$ case is similar.

\textit{Finite-cutoff holography---}A CFT defined at the boundary $\partial \mathcal{B}$ at infinity  is holographically dual to quantum gravity in the asymptotically AdS bulk region $\mathcal{B}$. Integrable and solvable deformations of two-dimensional CFTs with finite UV cutoffs have been extensively studied \cite{Zamolodchikov:2004ce,Smirnov:2016lqw}. The composite operator $T\bar{T}$, constructed from the stress-energy tensor $T_{\mu\nu}$ of the deformed quantum theory, exemplifies these irrelevant yet integrable deformations. Integrability guarantees that certain observables, such as the energy spectrum of the deformed theory, can be explicitly expressed in terms of the undeformed theory.

The generalization of $T\bar{T}$ deformations of two-dimensional theories with large central charge to higher-dimensional effective field theories with finite UV cutoffs and large degrees of freedom, dual to gravitational theories coupled with matter fields in the bulk, has been recently explored \cite{Hartman:2018tkw,Taylor:2018xcy}, along with extensions to the double-scaled SYK model case \cite{Aguilar-Gutierrez:2024oea}.  Unlike the two-dimensional scenario, higher-dimensional $T^2$ deformations explicitly involve background fields. The action $S_\lambda$ for a perturbed $d$-dimensional CFT is governed by an exactly solvable one-parameter flow equation
\begin{equation}\label{jfe9838}
\frac{\mathrm{d} S_\lambda}{\mathrm{d}\lambda} = \int \mathrm{d}^d x \sqrt{\gamma}\mathcal{F}(T_{ij}, G_{ij})
\end{equation}
where $S_{\lambda=0}$ is the undeformed theory, and $\mathcal{F}$ is a scalar function of the stress-energy tensor $T_{ij}$ and the background metric $G_{ij}$. The extent of the deformation is characterized by the double trace deformation parameter $\lambda > 0$. Negative values of $\lambda$ yield theories with a super-Hagedorn asymptotic density of states, resulting in ill-defined canonical ensembles \cite{Dubovsky:2012wk}, whereas    $\lambda >0$ corresponds holographically to an AdS bulk geometry with a finite radial cutoff, where the dual deformed CFT resides on a Dirichlet boundary. 

Physical observables such as the energy spectrum \cite{Smirnov:2016lqw,Cavaglia:2016oda,McGough:2016lol,Hartman:2018tkw}, stress-energy tensor correlation functions \cite{Kraus:2018xrn,Hartman:2018tkw,Cardy:2019qao,He:2023hoj,He:2023knl,He:2024xbi,He:2023kgq}, and entanglement entropy \cite{Donnelly:2018bef,Chen:2018eqk} are computable within this deformation framework. From the holographic viewpoint, a positively deformed CFT corresponds precisely to a bulk geometry truncated by a finite cutoff \cite{Heemskerk:2010hk,McGough:2016lol,Hartman:2018tkw,Taylor:2018xcy,Gross:2019ach,Caputa:2019pam,Jiang:2019epa,He:2025ppz}, described by the Fefferman–Graham expansion \cite{fefferman2012ambient,Graham:1999pm}
\begin{equation}\label{fje93822}
\mathrm{d}s^2 = \frac{\ell^2}{r^2}(\mathrm{d}r^2 + g_{ij}(x,r)\mathrm{d}x^i \mathrm{d}x^j)\,, \quad r < r_c
\end{equation}
where $\ell$ is the AdS radius, $x^i$ are a set of the boundary coordinates, and the metric $g_{ij}(x,r)=g_{(0) i j}+r^2 g_{(2) i j}+r^3 g_{(3) ij}+\cdots$ admits an expansion in powers of the radial coordinate $r$, which in turn yields the boundary stress-energy tensor \cite{deHaro:2000vlm,Skenderis:2002wp}. 

The holographic dictionary relating gravity and effective field theory is succinctly expressed as
\begin{equation}\label{fjoi34}
Z_{\mathrm{EFT}}[\lambda, \gamma_{ij}] = Z_{\mathrm{grav}}[g_{(0)}=r_c^2 \gamma_{ij}]
\end{equation}
with $\gamma_{ij}$ denoting the metric of the deformed CFT. This equivalence indicates that the generating functional of the deformed CFT defined by \eqref{jfe9838} matches the gravitational path integral for a finite cutoff geometry \eqref{fje93822}, applicable both to pure gravity \cite{McGough:2016lol} and to gravity theories coupled with matter \cite{Hartman:2018tkw}.

In what follows, we focus on the $d=3$ case. The standard holographic dictionary in four-dimensional spacetime is attained via \eqref{fjoi34} in the limit $r_c\to \infty$. The background metric of the original seed CFT in static coordinates is given by
\begin{equation}\label{jfqp394}
\mathrm{d} s^2=\omega^2(-\mathrm{d} t^2+\ell^2 \mathrm{d} \Omega_2^2)
\end{equation}
that can be obtained by a Weyl rescaling of the asymptotic AdS metric \cite{Emparan:1999pm,Emparan:1999gf}. Here $\omega$ is a variable conformal factor, and $\mathrm{d}\Omega_2^2$ is the line element of the unit sphere $S^2$. The thermodynamic first law 
\begin{equation}\label{jio34}
\mathrm{d} M=T \mathrm{d}S+V \mathrm{d} P+\cdots
\end{equation}
in the gravitational bulk is holographically dual to
\begin{equation}\label{jfgoi3}
\mathrm{d}\bar{E}=\bar{T} \mathrm{d} \bar{S}-\bar{P} \mathrm{d}\bar{V}+\mu \mathrm{d} c+\cdots
\end{equation}
on the conformal boundary at spatial infinity \cite{Ahmed:2023snm,Gong:2023ywu,Ahmed:2023dnh}. Here, $M, T, S, V, P$ respectively represent the bulk mass, temperature, entropy, thermodynamic volume, and thermodynamic pressure, while $\bar{E}, \bar{T}, \bar{S}, \bar{V}, \bar{P}$ denote their boundary counterparts. Note that there is an extra pair $(c,\mu)$ of conjugate variables,  with $c$ being the central charge  and $\mu$ its conjugate chemical potential, relative to the bulk. 
As pointed out in \cite{Visser:2021eqk}, the internal energy of the large-$N$ $S U(N)$ gauge theories with conformal symmetry depends not only on the extensive quantities $\bar{S}$ and $\bar{V}$, but also on the intensive quantity $c\sim N^2$. 
This is consistent with the scenario of working in the extended phase space, where $
P=3/(8 \pi \mathrm{G} \ell^2)
$
relates the AdS radius   to the thermodynamic pressure, with Newton's constant $\mathrm{G}$.  Variation of the pressure is equivalent to varying central charge and volume of the CFT. The standard extrapolated holographic dictionary reads \cite{Witten:1998zw,Gubser:1996de,Verlinde:2000wg,Savonije:2001nd}
\begin{equation}\label{jfi934j923}
M=\omega\bar{E},\quad S=\bar{S},\quad c=\frac{\ell^{2}}{4\mathrm{G}}\,.
\end{equation}
The holographic temperature relation is 
\begin{equation}
    T=\omega\bar{T}
\end{equation}
via the first laws for the bulk and boundary.

Turning to the finite-cutoff holographic thermodynamics, we now consider a deformed CFT residing at a finite radial location, anchored to a Dirichlet wall at $r = r_c$. To consistently implement  finite-cutoff holographic thermodynamics, the standard thermodynamic quantities must be suitably adapted. On the gravitational side, the bulk mass $M$ in \eqref{jio34} is replaced by a cutoff-dependent energy $\mathcal{E}(r_c)$, which approaches $M$ in the asymptotic limit as
\begin{equation}\label{fj3p9j39}
\lim_{r_c\to\infty}\mathcal{E}(r_c)= M\,.
\end{equation}
Correspondingly, the energy $\bar{E}$ of the undeformed boundary CFT in \eqref{jfgoi3} is generalized to a deformed energy $\bar{\mathcal{E}}(\lambda)$, which satisfies the UV consistency condition
\begin{equation}\label{jf934jefp}
\lim_{\lambda\to 0}\bar{\mathcal{E}}(\lambda) = \bar{E}\,.
\end{equation}
In light of \eqref{jfi934j923}, the new dictionary between the bulk mass and boundary deformed energies becomes
\begin{equation}\label{jfi349384}
\mathcal{E}(r_c) = \omega\bar{\mathcal{E}}(\lambda)
\end{equation}
since the conformal factor $\omega$ is independent of both the cutoff radius $r_c$ and the deformation parameter $\lambda$. Due to \eqref{fjoi34}, the holographic relation for the Bekenstein-Hawking entropy $S$ and the entropy $\bar{S}$ for the deformed CFT remains unchanged  \cite{McGough:2016lol}. Moreover, the Brown–Henneaux-like relation \cite{Brown:1986nw} in \eqref{jfi934j923} still holds.

We are now ready to construct   finite-cutoff holographic thermodynamics. Specifically, we consider a four-dimensional Schwarzschild-AdS (SAdS) black hole as the bulk configuration, subject to a finite Dirichlet cutoff. The finite-cutoff holographic duality between the $T^2$ deformation field theory and the cutoff SAdS geometry implies that this setup encodes the physics of a three-dimensional deformed CFT living on the finite cutoff surface.

\textit{Finite-cutoff holographic thermodynamics: boundary side---}The background geometry of the $T^2$ deformed CFT is specified by \cite{Hartman:2018tkw}
\begin{equation}
\mathrm{d} s^2 = -\mathrm{d} \tau^2 + R^2 \mathrm{d} \Omega_2^2
\end{equation}
where $R^2 \mathrm{d} \Omega_2^2$ denotes the standard metric on a two-sphere $S^2$ of radius $R$.  The spatial volume of the deformed CFT is thus
\begin{equation}\label{jfi3p94uio9}
\bar{V} = 4\pi R^2\,.
\end{equation}
This expression matches the volume inferred from \eqref{jfqp394} provided the conformal factor satisfies $\omega = R/\ell$. Note that in so doing, we have $\tau=\omega t$. The finite-size energy spectrum of a large-$N$ $T^2$ deformed three-dimensional CFT is given by \cite{Hartman:2018tkw}
\begin{equation}\label{fj3i94}
\bar{\mathcal{E}} = \frac{4 \pi R^2}{3 \lambda} \left(1 + \frac{3 \alpha \lambda^{2/3}}{R^2} - \sqrt{1 - \frac{3 \lambda \bar{E}}{2 \pi R^2} + \frac{6 \alpha \lambda^{2/3}}{R^2}}\right)
\end{equation}
where  $\alpha =\ell ^{4/3}/\left(4 \sqrt[3]{6} \pi ^{2/3} G^{2/3}\right)=c^{2/3}/\left(2 \sqrt[3]{3} \pi ^{2/3}\right)$, using the Brown–Henneaux-like relation \eqref{jfi934j923}  for $c$. In the UV limit $\lambda \to 0$,  \eqref{jf934jefp} is  satisfied,  with $\bar{E} $ representing the energy of the undeformed CFT;  explicitly 
\cite{Cong:2021jgb,Zhang:2023uay}
$
\bar{E} = \left( 4 \pi c \bar{S} + \bar{S}^2 \right) / \left( 2 \pi \sqrt{c \bar{S} \mathcal{V}_0} \right)
$.
Since the conformal factor $\omega$ can be chosen freely, we identify $\mathcal{V}_0 = \bar{V}$. Substituting this explicit form of $\bar{E}$ into \eqref{fj3i94} and considering \eqref{jfi3p94uio9}, we obtain the first law
\begin{equation}\label{bndry1law}
\mathrm{d}\bar{\mathcal{E}} = \bar{T} \mathrm{d} \bar{S} - \bar{P} \mathrm{d} \bar{V} + \mu \mathrm{d}c + \nu \mathrm{d}\lambda
\end{equation}
for the boundary of the cutoff bulk, 
where we have chemical potentials $\mu, \nu$  conjugate to the central charge $c$ and  the deformation parameter $\lambda$. All conjugate quantities in \eqref{bndry1law} can be directly attained via differentiation   using \eqref{fj3i94} (see Supplemental Material for details). The corresponding  holographic Euler relation for the deformed CFT reads
\begin{equation}
\bar{\mathcal{E}} = \bar{T} \bar{S} + \mu c - \nu \lambda\,.
\end{equation}
This relation reduces to the one for the seed CFT as $\lambda \to 0$ and it is a result of the scaling law
\begin{equation}
    \bar{\mathcal{E}}\left(\zeta \bar{S},\,\bar{V},\, \zeta c, \,\zeta^{-1}\lambda\right)=\zeta  \bar{\mathcal{E}}\left(\bar{S},\,\bar{V},\,  c, \,\lambda\right)
\end{equation}
where $\zeta$ is a dimensionless parameter. 

The equation of state for the seed CFT is $\bar{P}=\bar{E}/(2\bar{V})$ \cite{Ahmed:2023snm}. This must be modified for the  deformed CFT. We find another scaling law 
\begin{equation}
    \bar{\mathcal{E}}\left( \bar{S},\,\zeta^{2/3}\bar{V},\,  c, \,\zeta\lambda\right)=\zeta^{-1/3}  \bar{\mathcal{E}}\left(\bar{S},\,\bar{V},\,  c, \,\lambda\right)
\end{equation}
that yields the equation of state of the deformed CFT as
\begin{equation}
    \bar{P}=\frac{\bar{\mathcal{E}}}{2\bar{V}}+\frac{3\lambda\nu}{2\bar{V}}\,.
\end{equation}
According to \eqref{jf934jefp}, the standard equation of state of the seed CFT can be restored for $\lambda\to 0$.

\textit{Finite-cutoff holographic thermodynamics: bulk side---}The $T^2$ deformation of the three-dimensional CFT with large central charge is holographically dual to the SAdS black hole with a finite Dirichlet radial cutoff. In the framework of finite-cutoff holography, there is an agreement between the 
%square root 
dressed  energy spectrum \eqref{fj3i94} of the deformed field theory and the quasilocal energy of the SAdS black hole evaluated within the region $r < r_c$, given by \cite{Brown:1994gs}
\begin{equation}\label{12j09348jr}
\mathcal{M} = \frac{r_c}{\mathrm{G}}\left(\frac{\ell}{2 r_c} + \frac{r_c}{\ell} - \sqrt{1 - \frac{2 \mathrm{G} M}{r_c} + \frac{r_c^2}{\ell^2}}\right)
\end{equation}
where $M$ denotes the mass of the original SAdS black hole without a radial cutoff. Note that at large $r_c$, $\mathcal{M}\to M\ell/r_c+\mathcal{O}\left(1/r_c^2\right)$.
To recover the original mass $M$ and its associated first law in the absence of a cutoff, we define a rescaled bulk mass 
\begin{equation}\label{jfoei3993}
\mathcal{E} \equiv \frac{r_c \mathcal{M}}{\ell} = \frac{\sqrt{A_c} \left(-2 \sqrt{2} \sqrt{\mathbb{F} \mathrm{G} P}+4 \mathrm{G} P A_c+3\right)}{12 \sqrt{\pi } \mathrm{G}}
\end{equation}
where $\mathbb{F} =A_c \left(2 \mathrm{G} P A_c+3\right)-\sqrt{A_c}(2 \sqrt{\mathrm{G}S} (3 + 8 \mathrm{G}^2 P S))$ with $A_c \equiv 4 \pi r_c^2$ representing the size of the transverse space, i.e., the area of the cutoff surface. Here we have rescaled the quasilocal energy \eqref{12j09348jr} by a dimensionless factor $r_c/\ell$. The advantage of this redefined observable, which is invariant under all gauge transformations of the theory \cite{Marolf:2008tx}, is that for  $A_c \to \infty$, it reduces to the ADM mass $M$ of the SAdS black hole as $\lim_{A_c \to \infty} \mathcal{E}(A_c) \to M$. For the vacuum AdS case, it also reduces to zero energy. A similar rescaling was also used to calculate the holographic mass of   accelerating black holes  \cite{Anabalon:2018ydc}. Note that the boundary stress tensor $T_{i j}$ is related to the bulk Brown-York tensor $\widetilde{T}_{i j}$ via  $r_c\widetilde{T}_{i j}= T_{i j}$. The stripping off of the asymptotically  decaying factor $\ell/r_c$ in \eqref{jfoei3993} can be viewed as an approach to extracting the holographic mass of cutoff SAdS black hole,  similar to   methods  for extracting the full CFT stress-energy tensor   from the corresponding supergravity solutions \cite{Myers:1999psa}. Given that the mass of the SAdS black hole without a cutoff is $M =\sqrt{S}\left( 8 \mathrm{G}^2 P S +  3 \right)/ (6 \sqrt{\pi \mathrm{G}})$ \cite{Caldarelli:1999xj}, the first law of thermodynamics for the cutoff SAdS black hole takes the form
\begin{equation}\label{jf3928p4}
\mathrm{d}\mathcal{E} = T \mathrm{d}S + V \mathrm{d}P + \tau \mathrm{d}A_c\,,
\end{equation}
and the associated Smarr relation is
\begin{equation}
\mathcal{E} = 2 T S - 2 V P + 2 \tau A_c\,.
\end{equation}
The conjugate quantities, temperature $T$, thermodynamic volume $V$, and tension $\tau$, can be obtained via differentiation after substituting the bulk mass \eqref{jfoei3993}  into \eqref{jf3928p4} (see Supplemental Material for their explicit forms).  

To establish the relation in \eqref{jfi349384}, which connects the redefined energy $\bar{\mathcal{E}}$ of the deformed CFT in \eqref{fj3i94} to the modified bulk mass $\mathcal{E}$ in \eqref{jfoei3993} via a conformal scaling parameter $\omega$, we relate the cutoff radius $r_c$ to the deformation parameter $\lambda$ through the holographic dictionary
\begin{equation}\label{jfgi39e93}
\tilde{\lambda} \equiv \frac{\lambda}{R^3} = \frac{4 \pi \mathrm{G} \ell}{3 r_c^3}
\end{equation}
where $\tilde{\lambda}$ is the dimensionless reduced deformation parameter, which also allows the construction of bulk thermodynamics based on  $\bar{\mathcal{E}}(\bar{E}, \bar{V}, c, \tilde{\lambda})$. This dictionary makes $\ell\mathcal{E}=R \bar{\mathcal{E}}$; in the limit $r_c\to \infty, \lambda\to 0$, we have $\ell M=R \bar{E}$, consistent with \eqref{jfi934j923}.

The thermodynamic first law \eqref{bndry1law} for the deformed CFT and its bulk counterpart \eqref{jf3928p4}  can be derived from each other. To illustrate this connection, we outline how  the former can be obtained  from the latter. For the left side of \eqref{jf3928p4}, we have
$\mathrm{d}\mathcal{E}=\mathrm{d}\left(R \bar{\mathcal{E}}/\ell\right).$
For the right side of \eqref{jf3928p4}, expressing all the bulk quantities $S, P, A_c, \ell, r_c $ in terms of the boundary counterparts,  we just solve  $\mathrm{d}\bar{\mathcal{E}}$ to find that the $T\mathrm{d}S$ term becomes $\bar{T}\mathrm{d}\bar{S}$, the $V\mathrm{d}P$ terms along with part of $\tau \mathrm{d}A_c$ term become $-\bar{P}\mathrm{d}\bar{V}+\mu \mathrm{d}c$, and $\nu \mathrm{d}\lambda$ term arises from the remaining part of the $\tau \mathrm{d}A_c$ term (see Supplemental Material for details).

\begin{figure}[t!]
    \centering
     \includegraphics[width=8.5cm]{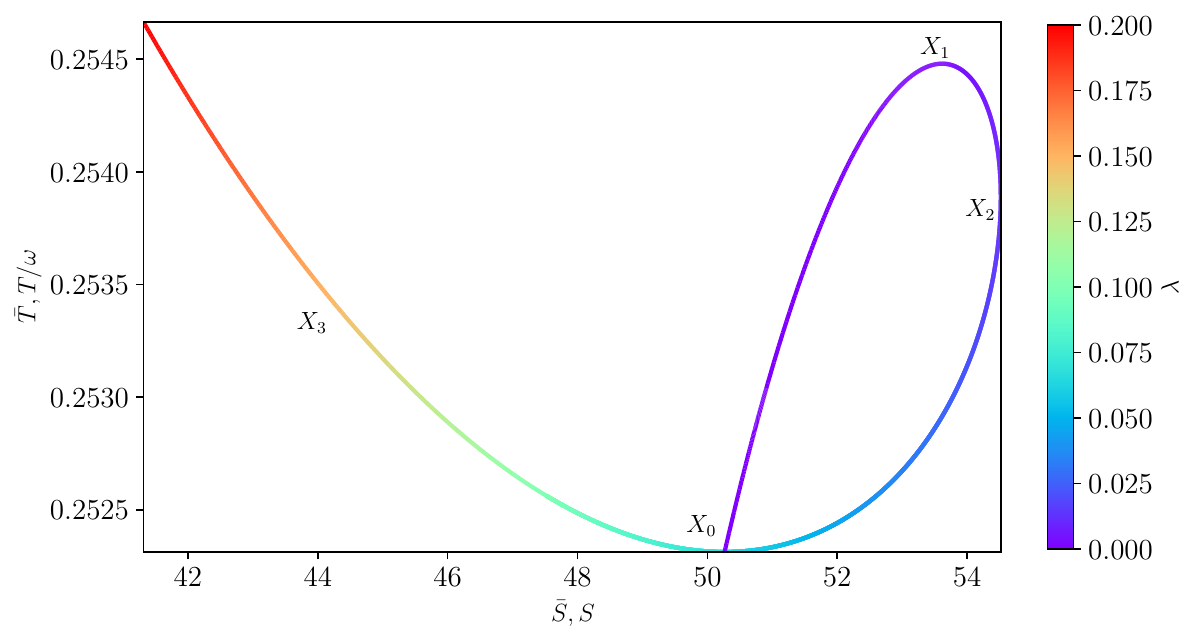} 
    \caption{Coexistence curve of the boundary confinement/deconfinement and bulk Hawking-Page phase transitions with $c=4, \bar{V}=20$. }
    \label{fjoie3k}
\end{figure}

\textit{Phase transitions---}In the cutoff SAdS bulk, the free energy is defined as $G = \mathcal{E} - T S$; setting $M = 0$  in \eqref{12j09348jr} yields that of the cutoff AdS vacuum. As in the standard SAdS scenario, a Hawking–Page phase transition occurs: as the temperature increases, the system transits from the cutoff vacuum AdS to the cutoff SAdS black hole. On the deformed CFT side, comparing the  free energy  given by $\bar{G} = \bar{\mathcal{E}} - \bar{T} \bar{S}$ with that of 
the deformed vacuum reveals a confinement/deconfinement phase transition. Figure~\ref{fjoie3k} shows the coexistence curve traced out by the phase transition points. Due to holographic duality, the $\bar{T}$ vs. $\bar{S}$ and $T/\omega$ vs. $S$ curves coincide and are plotted together.   Varying $A_c$ or $\lambda$  amounts to changing the UV cutoff scale of the deformed CFT and  selecting a family of $T^2$-deformed CFTs. This redefines the corresponding holographic RG flow starting from $A_c$ and also redefines the RG flow of the deformed CFT starting from a new UV scale. Note the deformation flow here is not the traditional RG flow; rather  it defines a family of effective field theories without coarse-graining. The point  $X_0$ marks the confinement-deconfinement transition point in the undeformed CFT or the Hawking–Page phase transition point in SAdS without a cutoff. As the deformation parameter $\lambda$ increases (reducing the cut-off $A_c$), the curve flows from $X_0 \to X_1 \to X_2 \to X_3$, forming a teardrop-shaped Rupert coexistence curve.  The region above the  segment $X_0 X_1 X_2$  corresponds to the deconfinement (bulk black hole) phase, while the region below corresponds to the confinement (AdS vacuum) phase. As $\lambda$ increases further, the curve extends as entropy  decreases back to $X_0$. The segment $X_2 X_0$ separates the deconfinement (black hole) phase above from the confinement (AdS vacuum) phase below.  The loop $X_0 X_1 X_2 X_0$  encloses an island where the phase varies under the deformation flow, surrounded  by an exterior region of fixed phase. Within   the Rupert curve, up to three deformed CFTs or cut-off SAdS configurations may share the same phase transition temperature, with one deformed CFT (or cut-off SAdS) matching the seed CFT’s (or original SAdS’s) phase transition temperature and entropy. Note that according to \eqref{fj3i94}, a shock singularity \cite{Smirnov:2016lqw} emerges for $\lambda>\pi  \sqrt{c} \bar{V}^{3/2}/3 \bar{S}^{3/2}$.

\textit{Discussion---}We have developed a novel framework for holographic black hole thermodynamics in  the cutoff holography setting, which incorporates finite radial cutoffs into the AdS/CFT correspondence. By formulating the first laws and their integral forms for a SAdS black hole with a Dirichlet cutoff and its dual $T^2$ deformed CFT, we established a precise holographic dictionary in which  the deformation parameter is a thermodynamic variable. We found that the Rupert teardrop coexistence curve defines a phase space island where states evolve with deformation flow, allowing up to three deformed CFTs or cut-off SAdS configurations to share the same phase transition temperature, with one matching the seed CFT or original SAdS. This illustrates the thermodynamic characteristics in  finite-cutoff holographic thermodynamics.   These results offer new insights into gravitational thermodynamics with boundary constraints and open a pathway toward understanding quantum gravity in finite regions of spacetime.

Based on our results, several directions for future research can be pursued. Our analysis can be readily extended to holographic thermodynamics for a CFT in general dimensions, including $d=2$, where $T\bar{T}$ holographic thermodynamics can be constructed. The framework can also be applied to black holes with non-spherically symmetric horizons or charged black holes. Additionally, it is natural to explore the Penrose inequality \cite{Penrose:1973um} and the reverse isoperimetric inequality \cite{Cvetic:2010jb} within cutoff holographic thermodynamics. The current study in AdS backgrounds can be extended to other contexts, such as cosmological holography \cite{Strominger:2001gp,Alishahiha:2004md,Gorbenko:2018oov,Lewkowycz:2019xse,Flauger:2022hie,Silverstein:2024xnr,Aguilar-Gutierrez:2024nst}, Cauchy holography \cite{Araujo-Regado:2022gvw,Araujo-Regado:2022jpj,Khan:2023ljg,Soni:2024aop}, and black holes with Lifshitz and hyperscaling violating asymptotics \cite{Cong:2024pvs}. Furthermore, extensions to holographic thermodynamics with a negative deformation parameter $\lambda$ \cite{Guica:2019nzm} can be considered; the Hagedorn singularity that can be calculated via the torus partition function \cite{Gu:2025tpy} may arise and be related to the phase transition in the bulk. Investigating subleading orders in $1/N$ beyond the classical level presents an additional challenge for future work. Finally, the well-posedness of the Dirichlet boundary condition \cite{Liu:2024ymn,Liu:2025xij} in finite-cutoff holography, the case of the mixed \cite{Guica:2019nzm,He:2021bhj,He:2023xnb} and more general \cite{Parvizi:2025shq,Parvizi:2025wsg} boundary conditions,  and other relevant interpretations of black hole thermodynamics from holographic perfect fluid \cite{Mancilla:2024spp}, particularly for the finite-cutoff scenario \cite{Tian:2014goa},  warrant further study.

\textit{Acknowledgements}\kern0pt---We are grateful to Rob Myers, David Kubizňák, Robie Hennigar, Xiaoyi Liu, Sergio E. Aguilar-Gutierrez, Song He, Yuan Sun, and Yunfeng Jiang for helpful discussions. This work was supported by the Natural Sciences and Engineering Research Council of Canada and the National Natural Science Foundation of China (Grant No. 12365010). M. Z., W. T., and M. L. were  supported by the China Scholarship Council Scholarship.  Research at Perimeter Institute is supported in part by the Government of Canada through the Department of Innovation, Science and Economic Development and by the Province of Ontario through the Ministry of Colleges and Universities.

\bibliography{refs}

\onecolumngrid

\newpage

\begin{center}  
{\Large\bf Supplemental Material for: Finite-cutoff Holographic Thermodynamics} 
\end{center}

\subsection{Explicit expressions for thermodynamic quantities}

We show explicit expressions for the thermodynamic quantities in the first laws for the $T^2$-deformed boundary field theory and the finite-cutoff Schwarzschild-AdS black hole in what follows. For the $T^2$-deformed CFT, the first law reads
\begin{equation}\label{fjeiop349jnr}
\mathrm{d} \overline{\mathcal{E}}=\bar{T} \mathrm{d} \bar{S}-\bar{P} \mathrm{d} \bar{V}+\mu \mathrm{d} c+\nu \mathrm{d} \lambda
\end{equation}
where the energy $\bar{\mathcal{E}}$, temperature $\bar{T}$, pressure $\bar{P}$, chemical potentials $\mu$, $\nu$ conjugate to entropy $\bar{S}$, volume $\bar{V}$, central charge $c$, and deformation parameter $\lambda$ respectively are
\begin{equation}
    \bar{\mathcal{E}}=\frac{\bar{V}}{3\lambda}\left(-\sqrt{\frac{2 c^{2/3} \lambda ^{2/3} K}{\bar{V}}-\frac{6 E \lambda }{\bar{V}}+1}+\frac{c^{2/3} \lambda ^{2/3} K}{\bar{V}}+1\right)\,,
\end{equation}
\begin{equation}
    \bar{T}=\left(\frac{\partial \bar{\mathcal{E}}}{\partial \bar{S}}\right)_{\bar{V},c,\lambda}=\frac{\bar{V} \bar{E}'\left(\bar{S}\right)}{\sqrt{\bar{V} \left(-6 \lambda  \bar{E}+\bar{V}+2 K (c \lambda )^{2/3}\right)}}\,,
\end{equation}
\begin{equation}
    \bar{P}=-\left(\frac{\partial \bar{\mathcal{E}}}{\partial \bar{V}}\right)_{\bar{S},c,\lambda}=-\frac{\bar{V} \left(3 \lambda  \bar{E}'\left(\bar{V}\right)-1\right)+3 \lambda  \bar{E}+\nu _0 \sqrt{\bar{V}}-K (c \lambda )^{2/3}}{3 \lambda  \nu _0 \sqrt{\bar{V}}}\,,
\end{equation}
\begin{equation}
    \mu =\left(\frac{\partial \bar{\mathcal{E}}}{
    \partial c }\right)_{\bar{S},\bar{V},\lambda}=\frac{\sqrt{\bar{V}} \left(9 \sqrt[3]{c \lambda } \bar{E}'(c)-2 K\right)+2 K \nu _0}{9 \nu _0 \sqrt[3]{c \lambda }}\,,
\end{equation}
\begin{equation}
    \nu =\left(\frac{\partial \bar{\mathcal{E}}}{
    \partial \lambda }\right)_{\bar{S},\bar{V},c}=\frac{\sqrt{\bar{V}} \left(4 K (c \lambda )^{2/3}-9 E \lambda \right)-3 \nu _0 \bar{V}+3 \bar{V}^{3/2}-K \nu _0 (c \lambda )^{2/3}}{9 \lambda ^2 \nu _0}
\end{equation}
where
\begin{equation}
    K=\frac{6 \pi }{\sqrt[3]{3} \pi ^{2/3}}\,,\quad \nu_0=\sqrt{\bar{V}+2 K (c \lambda )^{2/3}-6 E \lambda }\,,
\end{equation}
\begin{equation}
    E=\frac{4 \pi  c \bar{S}+\bar{S}^2}{2 \pi  \sqrt{c \bar{S} \bar{V}}}\,,\quad \bar{E}'\left(\bar{S}\right)=\frac{3 \bar{S}+4 \pi  c}{4 \pi  \sqrt{c \bar{S} \bar{V}}}\,,\quad \bar{E}'\left(\bar{V}\right)=-\frac{c \bar{S}^2 \left(\bar{S}+4 \pi  c\right)}{4 \pi  \left(c \bar{S} \bar{V}\right)^{3/2}}\,,\quad \bar{E}'\left(c\right)=\frac{\bar{S} \left(4 \pi  c-\bar{S}\right)}{4 \pi  c \sqrt{c \bar{S} \bar{V}}}\,.
\end{equation}
The first law in the finite-cutoff bulk is 
\begin{equation}\label{j93pjr4gf9p}
\mathrm{d} \mathcal{E}=T \mathrm{d} S+V \mathrm{d} P+\tau \mathrm{d} A_c
\end{equation}
where the quasilocal mass $\mathcal{E}$, temperature $T$, volume $V$, tension $\tau$ conjugate to entropy $S$, pressure $P$, and cutoff surface area $A_c$  respectively are
\begin{equation}
    \mathcal{E}=\frac{\sqrt{A_c} \left(4 P A_c-2 \nu _1+3\right)}{12 \sqrt{\pi }}\,,\quad  T=\left(\frac{\partial \mathcal{E}}{\partial S}\right)_{P,A_c}=\frac{2 P A_c M'(S)}{\nu _1}\,,
\end{equation}
\begin{equation}
    V=\left(\frac{\partial \mathcal{E}}{\partial P}\right)_{S,A_c}=\frac{A_c \left(\sqrt{A_c} \left(-4 P A_c+2 \nu _1-3\right)+12 \sqrt{\pi } \left(P M'(P)+M\right)\right)}{6 \sqrt{\pi } \nu _1}\,,
\end{equation}
\begin{equation}
    \tau=\left(\frac{\partial \mathcal{E}}{\partial A_c}\right)_{S,P}=\frac{24 \sqrt{\pi } M P \sqrt{A_c}-16 P^2 A_c^2+12 \left(\nu _1-1\right) P A_c+\nu _1 \left(3-2 \nu _1\right)}{24 \sqrt{\pi } \nu _1 \sqrt{A_c}}\,,
\end{equation}
with
\begin{equation}
    \nu_1=\sqrt{2 P A_c \left(2 P A_c+3\right)-24 \sqrt{\pi } M P \sqrt{A_c}}\,,\quad  M=\frac{\sqrt{S} (8 P S+3)}{6 \sqrt{\pi }}\,,\quad M'(S)=\frac{8 P S+1}{4 \sqrt{\pi } \sqrt{S}}\,,\quad M'(P)=\frac{4 S^{3/2}}{3 \sqrt{\pi }}\,.
\end{equation}
We have set   Newton's constant $\mathrm{G}=1$ for simplicity.

\subsection{Detailed derivation of the first law from the bulk to the boundary}

Here we show in detail how the first law \eqref{fjeiop349jnr} for the $T^2$ deformed CFT can be derived from first law \eqref{j93pjr4gf9p} for the Schwarzschild-AdS spacetime with finite-cutoff. The relations we will use include
\begin{align}\label{fkoje3i8rj9}
    \frac{\lambda}{R^3} = \frac{4 \pi \mathrm{G} \ell}{3 r_c^3}\,,\quad c =\frac{ \ell^2 }{ 4\mathrm{G}}\,,\quad \alpha = \frac{c^{2/3}}{2 \sqrt[3]{3} \pi ^{2/3}}\,,\quad P=\frac{3}{8 \pi \mathrm{G} \ell^2}\,,\quad \bar{V}=4\pi R^2\,.
\end{align}
On the left side of the bulk first law \eqref{j93pjr4gf9p}, we have 
\begin{equation}\label{jfe93p84p}
\mathrm{d}\mathcal{E}=\mathrm{d}\left(\frac{R\bar{\mathcal{E}}}{\ell}\right)=\frac{ \bar{\mathcal{E}}}{\ell}\mathrm{d} R+\frac{R}{\ell}\mathrm{d}\bar{\mathcal{E}}-\frac{ R \bar{\mathcal{E}}}{\ell^2}\mathrm{d}\ell\,.
\end{equation}
Applying \eqref{fkoje3i8rj9}, each term on the right side of \eqref{j93pjr4gf9p} turns to be
\begin{equation}\label{fjpi9384j93}
    T\mathrm{d}S=T\mathrm{d}\bar{S}\,,\quad V\mathrm{d}P=-\frac{3 V  }{4 \pi  G \ell ^3}\mathrm{d}\ell\,,\quad \tau \mathrm{d}A_c=8\pi r_c \tau \mathrm{d}r_c\,.
\end{equation}
Combing \eqref{j93pjr4gf9p}, \eqref{jfe93p84p}, and \eqref{fjpi9384j93}, we get
\begin{equation}\label{fj3984pi}
    \mathrm{d}\bar{\mathcal{E}}=-\frac{ \bar{\mathcal{E}}}{R}\mathrm{d}R+\frac{  \bar{\mathcal{E}}}{\ell}\mathrm{d}\ell+\frac{ T \ell }{R}\mathrm{d}\bar{S}+\frac{ 8 \pi  \tau  \ell  r_c}{R}\mathrm{d}r_c-\frac{3 V}{4 \pi  \mathrm{G} R \ell ^2}\mathrm{d}\ell\,.
\end{equation}
Furthermore, $\mathrm{d}r_c$ term can be split into
\begin{equation}\label{jyy}
    \frac{ 8 \pi  \tau  \ell  r_c}{R}\mathrm{d}r_c=\frac{16 \sqrt[3]{2} \pi ^{5/3} \mathrm{G}^{2/3} \ell^{2/3} \tau  \left(-\ell R \mathrm{d}\lambda+ \lambda  R \mathrm{d}\ell+3  \lambda  \ell \mathrm{d}R\right)}{ 3^{5/3} \lambda ^{5/3}}\,.
\end{equation}
Substituting \eqref{jyy} back into \eqref{fj3984pi} and then collect like terms, we have relations
\begin{equation}
    \frac{ T \ell }{R}\mathrm{d}\bar{S}=\bar{T}\mathrm{d}\bar{S}\,,\quad  \left(\frac{16 \sqrt[3]{2} \pi ^{5/3} \mathrm{G}^{2/3} \tau  \ell ^{5/3}}{3^{2/3} \lambda ^{2/3}}-\frac{\bar{\mathcal{E}}}{R}\right)\mathrm{d}R=-\bar{P} \mathrm{d} \bar{V}\,,
\end{equation}
\begin{equation}
   \left(\frac{\bar{\mathcal{E}}}{\ell }+\frac{16 \sqrt[3]{2} \pi ^{5/3} \mathrm{G}^{2/3} R \tau  \ell ^{2/3}}{ 3^{5/3} \lambda ^{2/3}}-\frac{3 V}{4 \pi  \mathrm{G} R \ell ^2}\right)\mathrm{d}\ell=\mu \mathrm{d}c\,,\quad  -\frac{16 \sqrt[3]{2} \pi ^{5/3} \mathrm{G}^{2/3} R \tau  \ell ^{5/3}}{ 3^{5/3} \lambda ^{5/3}}\mathrm{d}\lambda=\nu \mathrm{d}\lambda\,.
\end{equation}
Then \eqref{fj3984pi} will be reduced to the boundary first law \eqref{fjeiop349jnr}. Note that the holographic dictionary for the temperature 
$
    \bar{T}=T\ell/R=T/\omega
$
was used.

\end{document}